# Numerical Simulation of the Mechanical Properties of Carbon Nanotube Using the Atomistic-Continuum Mechanics

Chung-Jung Wu[1], Chan-Yen Chou[1], Cheng-Nan Han[1], and Kuo-Ning Chiang[2]

Advanced Microsystem Packaging and Nano-Mechanics Research Lab
Dept. of Power Mechanical Engineering
National Tsing Hua University
101, Sec. 2, Kuang-Fu Rd.,
Hsinchu, Taiwan 300, R.O.C.

**ABSTRACT**

This paper the utilizes atomistic-continuum mechanics (ACM) [1]-[3] to investigate the mechanical properties of single-walled carbon nanotubes (SWCNTs). By establishing a linkage between structural mechanics and molecular mechanics, not only the Young's moduli could be obtained but also the modal analysis could be achieved. In addition, according to atomistic-continuum mechanics and finite element method, an effective atomistic-continuum model is constructed to investigate the above two mechanical properties of SWCNTs with affordable computational time by personal computers. The validity of the results is demonstrated by comparing them with existing results.

## 1. INTRODUCTION

Much interest has been focused on the notable mechanical properties of carbon nanotubes, particularly in terms of their high Young's moduli. However, a large variation of Young's modulus was also disclosed for single-walled carbon nanotubes (SWCNTs) like 0.32-1.47 TPa by Salvetat et al. [4] and Yu et al.[5]. Considering the difficulties in the measurement of carbon nanotubes, the computer simulations based on reasonable physical models could provide experimentalists with guidance. The two main simulation models are the atomistic-based and the continuum-based models. The anterior is currently restricted within hundreds of atoms by recent computational technology. Several models are available for the continuum mechanics. Based on molecular dynamics (MD), Chang and Gao [6] proposed an analytical model to relate the elastic properties of SWCNT to its atomic structure. Zhang et al. [7] incorporated interatomic potentials into a continuum analysis without any parameter fitting to study the linear elastic modulus of a SWCNT. Li and Chou [8] considered a SWCNT as a frame, and used a beam element to simulate elastic moduli. However, the chemical bond is not allowed to bend [9], and the angle variation energy cannot be equated directly. In order to correlate the angle variation energy with continuum mechanics, Odegard et al. [10] presented a factitious rod in a truss model to study the effective geometry of a graphite sheet. Also, Leung et al. [11] proposed an idea on spatial periodic strain, and a truss model with factitious rod was established for the mechanics of zigzag SWCNTs.

Meanwhile, the vibrational properties of nanotubes have been studied by Krishnan et al. [12] and the amplitude of thermal vibrations of cantilevered nanotubes has been used for predicting their Young's modulus. But the potential of carbon nanotubes as high-frequency resonators has not been explored by experiments. Accordingly, Li and Chou [13] proposed a frame-like structure and simulated the vibrational behavior of carbon nanotubes by molecular-structural-mechanics method. The results showed that the fundamental frequencies of carbon nanotubes could reach the level of 10 GHz – 1.5 THz depending on the nanotube diameter and length.

In this research, an equivalent-spring structure is represented. A spring element is applied to transform the covalent bonds in the carbon nanotubes in order to describe the interatomic force between adjacent carbon atoms, and then the originally discrete atomic structure is analyzed in the continuum level. Moreover, in order to describe the bond-angle behavior of the covalent bonds, a factitious spring element between the carbon atoms along the two sides of each bond-angle is added. For the material properties of the spring element in the SWCNT structure, the Brenner's second-generation reactive empirical bond order (REBO) potential energy [14] is selected to describe the binding energy between carbon atoms, including both the bond-length and the bond-angle term. In addition, according to the atomistic-continuum mechanics and the finite element methods, the atomistic-continuum model is required to investigate Young's moduli of SWCNTs and the modal analysis is also achieved within an affordable computational time by means of personal computers.

---
[1] Research Assistant
[2] Professor, Corresponding Author

## 2. FUNDAMENTAL THEORY

### 2.1. Interatomic Potential For Carbon

Brenner [14] determined the interatomic potential for carbon atoms as represented by Eq. 1:

$$E_b = \sum_i \sum_{j(>i)} [V^R(r_{ij}) - b_{ij} V^A(r_{ij})] \quad (1)$$

For atoms i and j, $r_{ij}$ is the distance between atoms i and j, $V^R$ and $V^A$ are the respective repulsive and attractive pair terms given by Eq. 2 and 3, and $b_{ij}$ is a bond order between atoms i and j as represented by Eq. 4:

$$V^R(r) = f^c(r)\left(1 + \frac{Q}{r}\right) A e^{-\alpha r} \quad (2)$$

$$V^A(r) = f^c(r) \sum_{n=1,3} B_n e^{-\beta_n r} \quad (3)$$

The parameter Q, A, α, $B_n$, and $β_n$ are determined from the known physical properties of single (from diamond), conjugated double (from graphite), full double (from ethene), and triple (from ethyne) bonds. The function $f_c$ is a smooth cutoff function to limit the range of the potential.

$$b_{ij} = \frac{1}{2}\left[b_{ij}^{\sigma-\pi} + b_{ji}^{\sigma-\pi}\right] + b_{ij}^{\pi} \quad (4)$$

The values for the functions $b_{ij}^{\sigma-\pi}$ and $b_{ji}^{\sigma-\pi}$ depend on the local coordination and bond angles for atoms i and j, respectively. The function $b_{ij}^{\pi}$ is further written as a sum of two terms as given by Eq. 5:

$$b_{ij}^{\pi} = \Pi_{ij}^{RC} + b_{ij}^{DH} \quad (5)$$

The value of the first term $\Pi^{RC}_{ij}$ depends on whether a bond between atoms i and j has radical character and is part of a conjugated system. The value of the second term $b^{DH}_{ij}$ depends on the dihedral angle for carbon–carbon double bonds.

According to the REBO potential for solid carbon, the interatomic potential function including bond stretching, angle variation, and dihedral term can be obtained. Moreover, van der Waals and electrostatic interactions between carbon atoms can be derived from Lennard-Jones "6-12" potential and typical electrostatic potential, respectively. However, for a SWCNT that is subjected to axial loadings at small strains, dihedral, van der Waals, and the electrostatic potential could be negligible. Only bond stretching and angle variation terms are significant in the system potential energy.

### 2.2. Atomistic-Continuum Method

Based on the finite element method, the atomistic–continuum mechanics is developed to simulate the mechanical characteristics, such as the Young's modulus and Poisson's ratio of nanoscale structures. The ACM method transfers an originally discrete atomic potential into an equilibrium continuum model by atomistic-continuum transfer elements. It simplifies the complexities of the interactive forces among the atoms, while keeping the calculation accuracy still acceptable and the computational time affordable.

One can generate the equations for a typical static constant-strain finite element. The total potential energy is a function of the nodal displacements x, y and z such that $π_p = π_p(x,y,z)$. Here the total potential energy is given by Eq. 6

$$\pi_p = U + \Omega_b + \Omega_p + \Omega_s \quad (6)$$

where U, $\Omega_b$, $\Omega_p$ and $\Omega_s$ represent the strain energy, the potential energy of the body force, the potential energy of the concentrated load and the potential energy of the distributed load, respectively. The above equation can be rewritten as a finite element integrated form as Eq. 7 shown:

$$\pi_p = \frac{1}{2}\iiint_V \{d\}^T[B]^T[D][B]\{d\}dV - \iiint_V \{d\}^T[N]^T\{F\}dV - \{d\}^T\{P\} - \iint_S \{d\}^T[N_s]^T\{T_s\}dS \quad (7)$$

where {d} represents the nodal displacement vector, [B] is the strain-displacement matrix, [D] is the modulus of the elasticity matrix, [N] is the shape function matrix, {F} is the body force vector, {P} is the external load vector and {$T_s$} is the traction force vector.

The ACM method transfers the interatomic potential function into a force-displacement curve so as to create an equivalent atomistic-continuum transfer element. Afterwards, the equivalent nanoscale model can be analyzed by FEM.

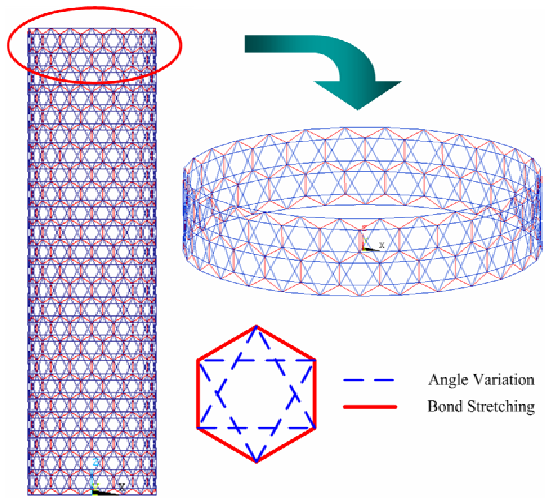

Figure 1. The equivalent-spring structure of a zigzag single-walled carbon nanotube.

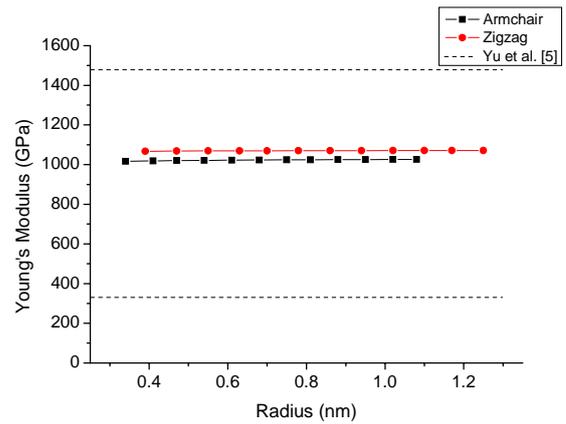

Figure 2. The result of Young's modulus vs. radius for a zigzag-type and armchair-type SWCNT.

### 2.3. SWCNT ACM Numerical Modeling

In this section, an ACM model was constructed to simulate the Young's modulus of SWCNT. A single-walled carbon nanotube, which can be viewed as a graphene sheet rolled into a tube, is usually indexed by a pair of integers (n, m) to represent its helicity. Based on the geometry described by the integers (n, m), the positions of carbon atoms in the SWCNT ACM model could be obtained.

A typical spring element is chosen as the equivalent atomistic-continuum transfer element. The distinct characteristics of a spring element from a truss or a beam element are that it is not allowed to bend, no cross-sectional area needs to be defined, and the potential between atoms is the same as spring element. According to the former two characteristics, a spring element could represent a more realistic equivalent model since the chemical bond could neither be bent nor be defined by a cross-sectional area. Moreover, based on the present potential theory, the last feature makes the transformation between the atomistic and the continuum method easier and more direct.

In this SWCNT ACM model, the chemical bond between carbon atoms could be transformed into the spring element. The carbon atoms can be viewed as the linkage between spring elements, and can be transformed into the pin-joint which cannot resist any bending moment. In addition, the factitious springs are needed either to equate the angle variation potential or to make the equivalent-spring model stable. Fig. 1 shows the SWCNT ACM model. The hexagon is a representative structure of SWCNT. The solid line in the perimeter of the hexagon and the dashed line in the hexagon represent the bond stretching and angle variation, respectively. One end of the SWCNT was fixed, and the other was applied with strain.

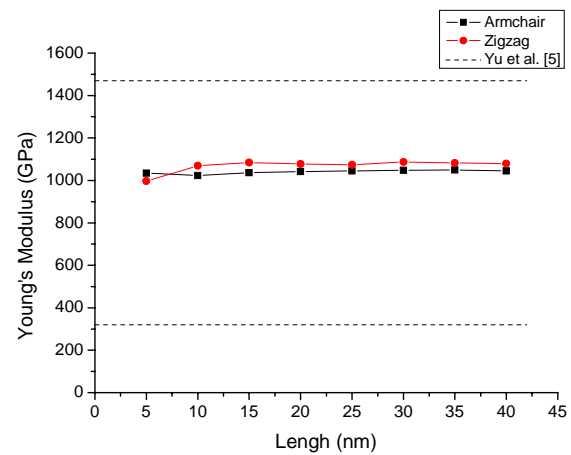

Figure 3. The result of Young's modulus vs. length for a zigzag-type and armchair-type SWCNT.

Meanwhile, by considering the atomic structure of carbon nanotube, the masses of electrons are neglected and the masses of carbon nuclei ($m_c = 1.99 \times 10^{-26}$) are assumed to be concentrated at the centers of atoms, which are equivalent to the mass of nodes in the ACM model as the modal analysis is proceeded.

### 3. SIMULATION RESULTS

### 3.1. Young's Modulus of SWCNT

Two geometric variables, namely, radius and length, with respect to zigzag-type and armchair-type single-walled carbon nanotubes, were analyzed in this research. In the case of the radius-variable, the length of SWCNT is chosen as 5 nm. Moreover, in the case of the length-variable, the integers of SWCNT are selected as (10, 10) and (18, 0) for armchair-type and zigzag-type SWCNT, respectively. The result of Young's modulus vs. the radius for zigzag and armchair SWCNT is shown in Fig. 2. The result of Young's modulus vs. the length for zigzag and armchair SWCNT is shown in Fig. 3. Computations on the elastic deformation of SWCNT reveal that the Young's moduli of both zigzag and armchair carbon nanotubes remain constant within a specific range of both 0.6~2.5 nm (radius) and 5~40 nm (length).

Table 1. Modal analysis results of SWCNT. Those shadowed and underlined mode frequency represented the mode shape was shell-mode while the else stood for beam-mode.

| $C_h$ (n,n) | D (nm) | L (nm) | Mode frequency (GHz) | | | | |
|---|---|---|---|---|---|---|---|
| | | | 1 | 2 | 3 | 4 | 5 |
| (8,8) | 1.1 | 80 | 4.752 | 13.045 | 25.437 | 41.771 | 61.912 |
| | | 90 | 3.746 | 10.294 | 20.094 | 33.042 | 49.052 |
| | | 100 | 3.037 | 8.348 | 16.309 | 26.845 | 39.898 |
| (15,15) | 2.2 | 80 | 8.842 | _12.397_ | _15.510_ | _21.596_ | 24.033 |
| | | 90 | 6.983 | _12.166_ | _14.409_ | _18.901_ | 19.036 |
| | | 100 | 5.667 | _12.019_ | _13.699_ | 15.481 | _17.105_ |
| (14,0) | 1.1 | 80 | 4.723 | 12.968 | 25.29 | 41.537 | 61.579 |
| | | 90 | 3.756 | 10.322 | 20.151 | 33.141 | 49.207 |
| | | 100 | 3.037 | 8.351 | 16.316 | 26.858 | 39.922 |
| (26,0) | 2.2 | 80 | 8.756 | _20.090_ | 23.807 | _23.995_ | _30.538_ |
| | | 90 | 6.976 | 19.019 | _19.795_ | _22.814_ | _27.905_ |
| | | 100 | 5.648 | 15.429 | _19.584_ | 21.965 | _25.994_ |

The Young's modulus is about 1,000 GPa, which falls within the range of Young's modulus as reported by Cornwell and Wille [15] (MD), Salvetat et al.[4], Yu et al. [4] (experimental result), and Zhang et al. [7] (continuum theory). However, the result is lower than other reported Young's moduli using ACM with a truss or a beam element (e.g., Li and Chou [8]; Odegard et al.[10]; Leung et al. [11]).

The results also show that the Young's moduli of armchair and zigzag SWCNT remain constant within the specific range in this research in terms of both radius and length. For a given tube radius, the Young's modulus for armchair SWCNT is slightly larger than that for zigzag SWCNT. These two results agree with the physical phenomena reported by Chang and Gao [6] even if the parameter of the tube length is different. In addition, for a given tube length, the Young's modulus for armchair SWCNT is slightly smaller than that for zigzag SWCNT beside the 5 nm tube length.

### 3.2. Modal Analysis of SWCNT

In this section, modal analysis of SWCNT was proceeded. The radius and length of SWCNT were chosen again as the variables with respect to the armchair-type and zigzag-type SWCNT, and the detail parameters were shown in the table 1. The parameters were selected to make the length-to-radius ratio of SWCNT large enough to ensure that the first mode could be a beam-mode instead of a shell-mode. The results of the first five mode frequency were listed in the table 1 and those shadowed and underlined mode frequency represented the mode shape was shell-mode while the else stood for beam-mode. The mode shape of beam-mode and shell-mode was shown in the Fig. 4.

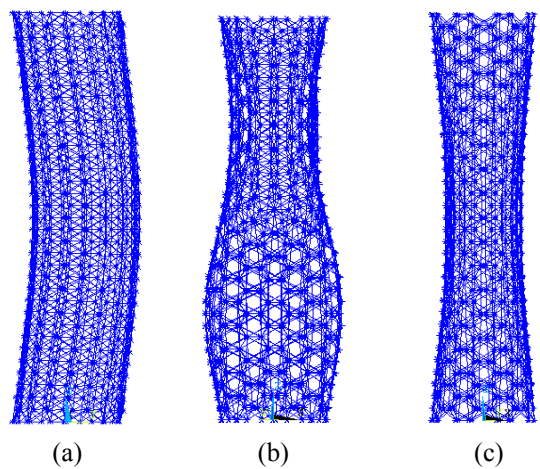

Figure 4. The representative mode shape of SWCNT ACM model. The mode shape of beam-mode was shown as (a) while (b) and (c) were shell-mode.

By comparing the mode frequency with the same mode shape, the results showed that the mode frequency was proportional to the radius, but inverse proportional to the length. The results agreed with vibrational properties of the Bernoulli-Euler beam derived by the classical structural mechanics [16] and the simulation results published by Li and Chou [13].

### 4. CONCLUSION

In this research, an atomistic-continuum mechanical (ACM) model based on the finite element method with an equivalent-spring element was proposed to investigation the Young's modulus and the modal analysis of SWCNT. The spring element could provide a more realistic equivalent model between atomistic and continuum mechanics. The result agreed with some experimental and atomistic studies based on ab initio, MD, and the continuum theory. Moreover, the present approach is not only limited to SWCNT since other atomistic studies that provide potential in atomic bonds can be similarly incorporated in the present ACM modes with this equivalent-spring element.

## 5. ACKNOWLEDGEMENT

The authors would like to thank the **National Center for High-performance Computing (NCHC)** for supporting this research and **the National Science Council of Taiwan R.O.C.** for financially supporting this research under contract No. NSC 93-2120-M-007-005.